# Graphene – Based Nanocomposites as Highly Efficient Thermal Interface Materials

## Khan M. F. Shahil[×] and Alexander A. Balandin[*]

*Nano-Device Laboratory, Department of Electrical Engineering and Materials Science and Engineering Program, Bourns College of Engineering, University of California – Riverside, Riverside, California, 92521 USA*

## Abstract

We found that an optimized mixture of graphene and multilayer graphene – produced by the high-yield inexpensive liquid-phase-exfoliation technique – can lead to an extremely strong enhancement of the cross-plane thermal conductivity $K$ of the composite. The "laser flash" measurements revealed a record-high enhancement of $K$ by 2300 % in the graphene-based polymer at the filler loading fraction $f$ =10 vol. %. It was determined that a relatively high concentration of single-layer and bilayer graphene flakes (~10-15%) present simultaneously with thicker multilayers of large lateral size (~ 1 μm) were essential for the observed unusual $K$ enhancement. The thermal conductivity of a commercial thermal grease was increased from an initial value of ~5.8 W/mK to $K$=14 W/mK at the small loading $f$=2%, which preserved all mechanical properties of the hybrid. Our modeling results suggest that graphene – multilayer graphene nanocomposite used as the thermal interface material outperforms those with carbon nanotubes or metal nanoparticles owing to graphene's aspect ratio and lower Kapitza resistance at the graphene - matrix interface.

*Corresponding author (AAB): balandin@ee.ucr.edu ; http://ndl.ee.ucr.edu/

Rapidly increasing power densities in electronics made efficient heat removal a crucial issue for progress in information, communication and energy storage technologies [1-2]. Development of the next generations of integrated circuits (ICs), three-dimensional (3D) integration and ultra-fast high-power density communication devices makes the thermal management requirements extremely severe [1-6]. Efficient heat removal became a critical issue for the performance and reliability of modern electronic, optoelectronic, photonic devices and systems. Thermal interface materials (TIMs), applied between heat sources and heat sinks, are essential ingredients of thermal management [2-6]. Conventional TIMs filled with thermally conductive particles require high volume fractions $f$ of filler ($f\sim50\%$) to achieve thermal conductivity $K$ of the composite in the range of $\sim$1-5 W/mK at room temperature (RT) [3-6]. Earlier attempts of utilizing highly thermally conductive nanomaterials, e.g. carbon nanotubes (CNTs), as fillers in TIMs, have not led to practical applications due to weak thermal coupling at CNTs/base interface and prohibitive cost.

In this letter, we show that a proper mixture of graphene [7] and multilayer graphene (MLG) produced by high-yield liquid-phase-exfoliation (LPE) technique [8-10] can be used for TIMs with the strongly enhanced cross-plane (through-plane) thermal conductivity $K$. Moreover, it is demonstrated that our approach allows one to significantly improve the heat conduction properties of the commercial thermal greases with a very small addition ($f\sim2\%$) of the graphene-MGL filler. Experiments and simulations suggest that more efficient TIMs, which are used to minimize the thermal resistance between two surfaces (see Figure 1), can help to significantly lower the average and hot-spot temperatures in ICs. Achieving enhancement of TIMs' thermal conductivity by a factor of 10-20 compared to that of the matrix materials would revolutionize not only electronics but also renewable energy generation where temperature rise in solar cells degrades the performance and limits life-time.

TIM's function is to fill the voids and grooves created by imperfect surface finish of mating surfaces. Their performance is characterized by $R_{TIM}=BLT/K+R_{C1}+R_{C2}$, where $BLT$ is the bond line thickness and $R_{C1,2}$ are the TIM's contact resistance with the two bounding surfaces. The magnitude of $R_{TIM}$ depends on the surface roughness, interface pressure $P$, temperature $T$, and viscosity $\xi$. The common TIMs are composites, which consist of polymer matrix or base material and thermally conductive filler particles. TIMs have to be mechanical stabile, reliable, non-toxic, low-cost and easy to apply [2-6]. They should possess as high $K$ as possible, as well as low $\xi$ and coefficient of thermal expansion. The industrial TIMs have $R_{TIM}$ ~3-10×10⁻⁶ Km²/W [3]. The drive to reduce $L$ of the conventional fillers, e.g. metal particles, is explained by the fact that smaller $L$ at high $f$ results in larger particle-to-particle contact area and lower $R_{TIM}$ [6]. The efficiency of the filler in TIMs is characterized by the thermal conductivity enhancement (TCE) defined as $\eta = (K - K_m)/K_m$, where $K$ is thermal conductivity of the composite and $K_m$ is thermal conductivity of the matrix material. TCE of ~170% at the 50% loading of conventional fillers such as silver or alumina with the filler particle size $L<10$ μm are considered to be standard.

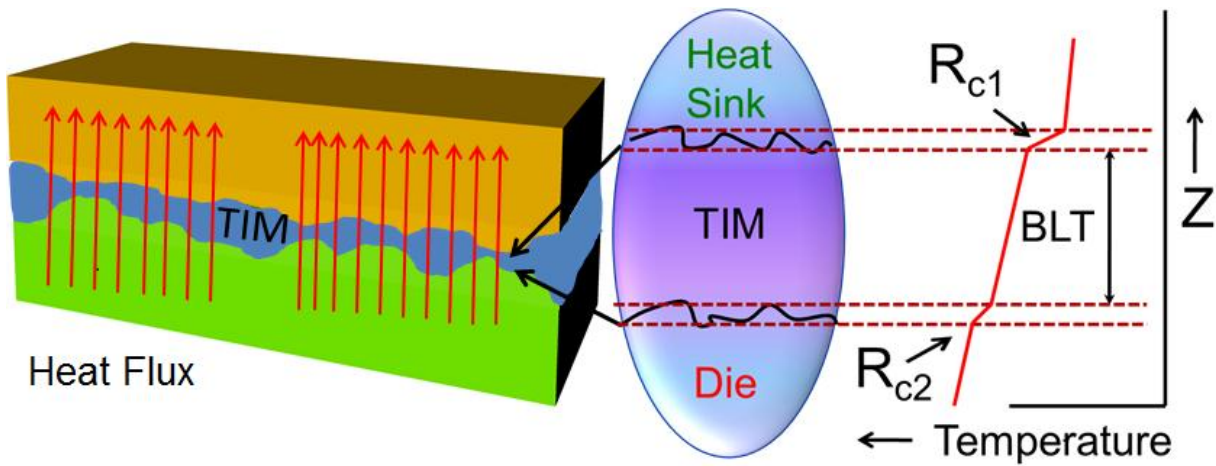

**Figure 1**: Schematic illustrating the action of thermal interface material, which fills the gaps between two contacting surfaces. The heat removal improves with higher thermal conductivity, smaller bond line thickness and contact resistance of the material.

A decade ago, CNTs attracted attention as potential fillers for TIMs. Their main attractive feature is extremely high intrinsic thermal conductivity $K_i$ of CNTs in the range of ~3000–3500 W/mK at RT [11-12]. The outcomes of experiments with CNT-based TIMs were controversial. The measured TCE factors were moderate, in the range ~50-250% at $f$~7% of the CNT loading [13-15]. In some cases, $K$ was not improved substantially [14] or even decreased with addition of single-wall CNTs [16]. The common reason offered as an explanation was that although CNTs have excellent $K_i$ they do not couple well to the matrix material or contact surface. The reported thermal boundary resistance (TBR) between CNTs and polymer matrix was as high as ~$10^{-7}$ m$^2$KW$^{-1}$ [17]. The large TBR at CNT/matrix interface can be attributed to the fundamental property – high Kapitza resistance [18] between one-dimensional (1D) CNTs and 3D bulk owing to the large difference in the phonon density of states (DOS). It was also suggested that the lack of *thermal percolation* in CNT composites can negatively affect their heat conduction properties [19]. Interestingly, the electrical percolation thresholds $f_T$ for CNT composites are very low, $f$~0.1 vol. %, compared to 20–30 vol. % for composites with spherical fillers [6, 19]. TIMs with aligned CNTs have better $K$ but suffer from large $R_C$ and prohibitive cost. These outcomes provide strong motivation for the search of alternative high-$K$ fillers.

Recently, it was discovered that graphene has extremely high intrinsic $K_i$, which exceeds that of CNTs [20-23]. The latter was confirmed by theoretical studies [21-23]. MLG retains good thermal properties [22-23]. Graphite, which is 3D bulk limit for MLG with the number of layers $n\rightarrow\infty$, is still an excellent heat conductor with $K_i\approx2000$ W/mK at RT. For comparison, $K_i\approx430$ W/mK for silver and it is much lower for silver nanoparticles used in TIMs. To test graphene as the TIM filler we adopted the surfactant stabilized graphene dispersion method [8-10] and graphene composite preparation techniques [24-25], with several modifications for maximizing TCE. The chosen approach requires relatively little chemical and thermal treatment and allows one to produce sufficient quantities of TIMs for detail study (Figure 2). The dispersions were prepared by ultrasonication of graphite flakes in aqueous solution of sodium cholate followed by sonication and centrifugation (see Methods section for details). In the extensive trial-and-error procedure we determined the optimum sonication time $t_s$ and centrifugation rate $r_c$ resulting in the largest $\eta$. Major advantages of the employed technique are the use of readily available graphite, low cost and scalability of production.

Figure 2 presents the optical, scanning electron microscopy (SEM) and atomic force microscopy (AFM) data for the synthesized material. The thickness $H$ (=$h\times n$, where $h$=0.35 nm is the thickness of graphene monolayer) and size $L$ distribution of MLG in the nanocomposites were important for maximizing $\eta$. We refer to the synthesized materials as nanocomposites because substantial portion of the filler particles had at least one dimension (thickness) below a few nanometers in size and the presence of these nanoscale components was essential for the materials' functionality.

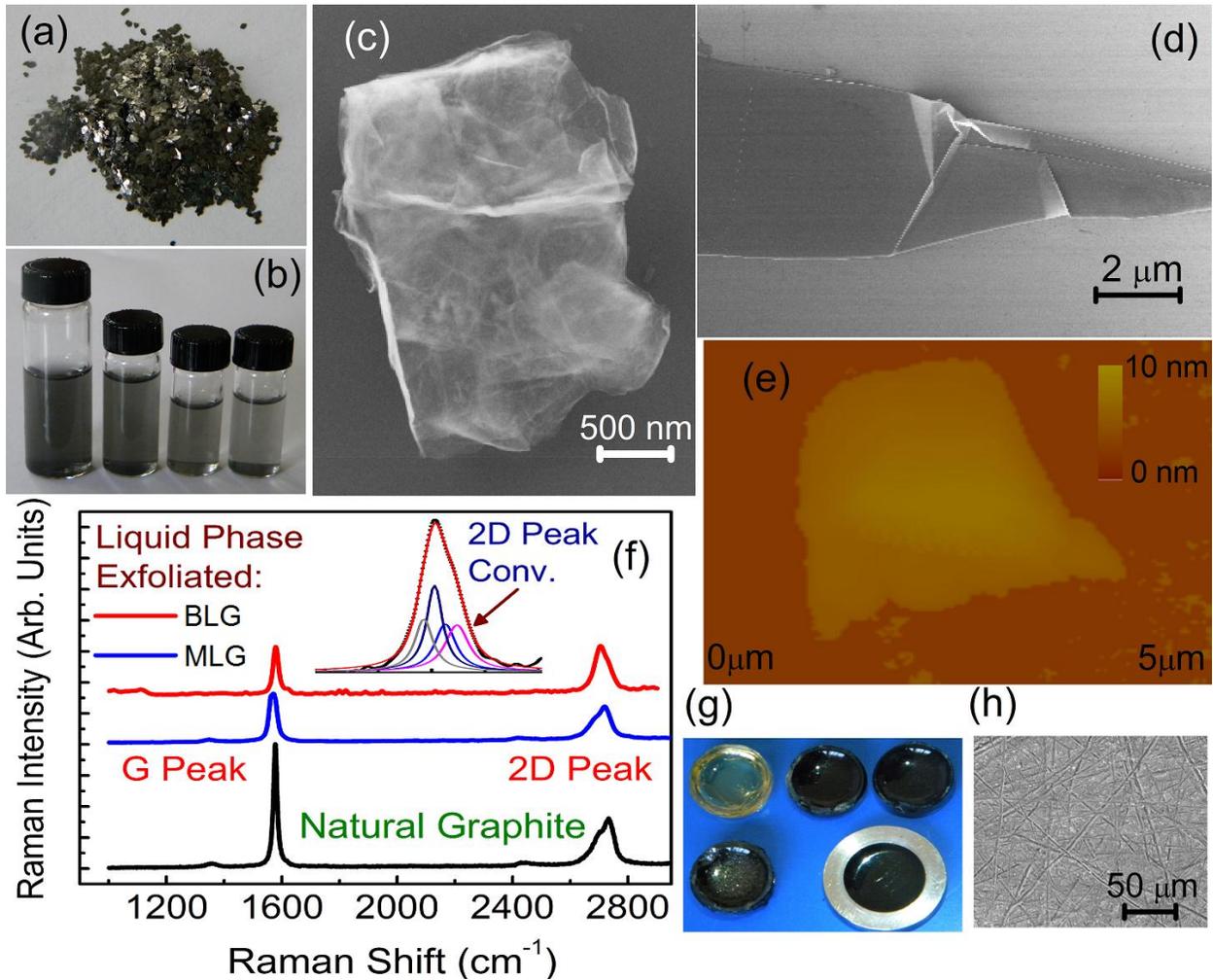

**Figure 2:** Synthesis and characterization of the graphene-MLG polymer nanocomposite TIMs. (a) graphite source material; (b) liquid-phase exfoliated graphene and MLG in solution; (c) SEM image of MLG revealing overlapping regions and wrinkles, which improve thermal coupling. (d) SEM image of a large MLG (*n*<5) flake extracted from the solution; (e) AFM image of MLG flake with varying *n*; (f) Raman spectroscopy image of bilayer graphene flakes extracted from the solution; (g) optical image of graphene-MLG polymer composite samples prepared for thermal measurements; (d) representative SEM image of the surface of the resulting graphene based TIMs indicating small roughness and excellent uniformity of the dispersion.

We used micro-Raman spectroscopy to verify $n$ [26]. The $n$ counting with the Raman spectroscopy is efficient for $n<7$. For thicker flakes the thickness distribution statistics was also derived from AFM inspection. Figure 2 shows an example of Raman spectra of MLG from the solution and the reference graphite source excited at $\lambda$=488 nm. Deconvolution of $2D$ band and comparison of the $I(G)/I(2D)$ intensity ratio allowed us to determine $n$ with a high accuracy, e.g. the plotted spectra correspond to the large-size bilayer graphene ($n$=2) and MLG with $n\approx5$ with. The weak intensity of the disorder $D$ peak, composed of the $A_{1g}$ zone-edge phonons, indicates the large size and low defect concentration. The diameter of the laser spot in the micro-Raman spectroscopy was ~1 μm. The graphene-MLG concentrations utilized for preparation of the nanocomposite TIMs were ~0.253 mg/mL ($t_s\approx$12 hrs, $r_c$=15 K-rpm). From the statistical analysis we established that the composites with ~10-15% of MLG with $n\leq2$, ~50% of FLG with $n\leq5$ are the optimum for maximizing $\eta$. Based on the optical microscopy and SEM examination, most of the graphene and MLG flakes (~90%) had lateral dimensions in the range $L\approx$50 nm - 0.5 μm. A small fraction of the flakes (~10%), predominantly with $n<5$, had large lateral sizes $L\approx$2-5 μm. As discussed below, their presence in the composites was important. The prepared nanocomposite graphene-MLG solutions were mixed with epoxy followed by curing and heating in vacuum to produce a large number of samples with the carbon loading $f$ ~0.2-10vol.%. The homogeneity of the resulting composite and adhesive bonding have been verified with SEM. It is important to mention here that our graphene-MLG fillers were substantially different from what is referred to as graphite nanoplatelets (GnP) characterized by higher thickness ($10-100$ nm).

Thermal conductivity measurements were conducted with the "laser flash" technique (NETZSCH). In the laser flash method, for a given geometry of the samples, heat propagates from the top to the bottom surface of the material under test. It means that the measured K is more closely related to the cross-plane (through plane) component of the thermal conductivity tensor. The cross-plane K is the one, which is important for TIMs' performance. The sample thicknesses were 1-1.5 mm to ensure that their thermal resistances were much larger than the contact resistance. As a control experiment we measured thermal conductivity of pristine epoxy and obtained $K$=0.201 W/mK at RT, in agreement with the epoxy vendor's specifications.

Figure 3a shows TCE factor as a function of $f$ for the graphene-MLG-nanocomposite epoxy - sample A ($t_s\approx$12 hrs, $r_c$=15 K-rpm) and sample B ($t_s\approx$10 hrs, $r_c$= 5 K-rpm) at RT. For comparison, we also measured TCE for the epoxy composites with graphite micro- and nanoparticles obtained by grinding the same graphite (substantial fraction of particles have $L\sim$40 μm) and for the epoxy composite with commercial carbon black (CB) powder. One can see that there is extraordinary increase in $\eta$ for our graphene-MLG nanocomposites. At $f$=10 vol. % loading, $K$ reaches the value of ~5.1 W/mK, which corresponds to TCE of ~2300 %. Traditional fillers with small aspect ratios show TCE ~20% per 1 vol. % loading. The measured TCE for composites with the amorphous graphite particles were low and consistent with the literature [2-6]. There were almost no TCE in CB-epoxy composites for the examined loading fractions. The control experiments with graphite particles and CB confirm that thermal properties of graphene and MLG were essential for increasing $K$ of TIMs.

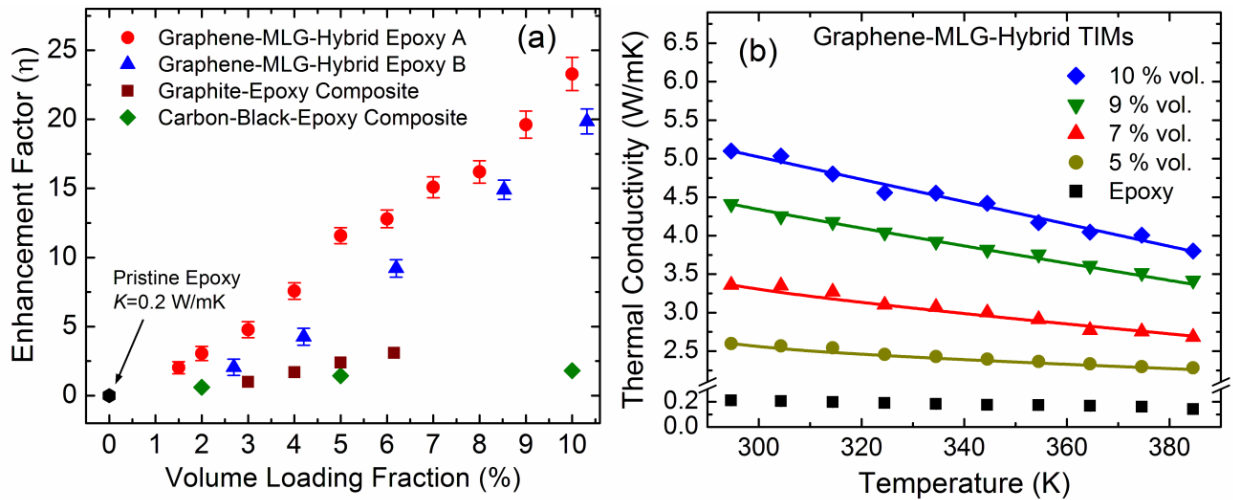

**Figure 3:** Thermal conductivity enhancement in the graphene-MLG-nanocomposite epoxy TIMs. (a) Measured thermal conductivity enhancement factor as a fraction of the filler volume loading fraction. Note an extremely large enhancement of ~2300% at $f$=10 vol. % for the optimum nanocomposite. (b) Experimentally determined dependence of thermal conductivity of TIMs on temperature for different loading fractions.

It is interesting to note that TCE follows approximately linear dependence on $f$ without revealing any clear signature of thermal percolation threshold. One would expect to observe a kink in $K(f)$ plot and $K = K_T[(f - f_T)/(1 - f_T)]^{\beta}$ dependence (where $\beta\approx$2 in 3D) if the percolation is resolved ($f$ step is 1 vol. % in our measurements). The physics of thermal percolation is still a subject of intense debates [2, 14, 19, 24]. Unlike electrical percolation the thermal percolation threshold can be less pronounced due to heat conduction by the matrix. Our attempts to increase $f$ of the graphene-MLG-polymer composites beyond 10 vol. % while maintaining acceptable TIM characteristics, e.g. $\xi$, $R_C$, were not successful. The changes in $\xi$ lead to inhomogeneous inclusions and surface roughness for $f>10$ vol. %. Figure 3b shows $K$ as a function of $T$ for different $f$. The $K$ decrease with $T$ at higher loading is reminiscent of the Umklapp phonon scattering characteristic for crystalline materials, including graphene [23]. This suggests that heat is carried by the thermally linked graphene or MLG flakes when $f\sim10\%$. Contrary, in pristine epoxy the $K(T)$ dependence is nearly absent, which is expected for the non-crystalline amorphous solids [23].

To analyze our experimental data, we used the Maxwell-Garnett effective medium approximation (EMA), which works well for $f<40\%$ [27-28]. We modified it to include the size of the fillers, aspect ratios $\alpha$ and TBR between the fillers and matrix (Supplemental Information). Both graphene and CNTs can be regarded as spheroids with principle dimensions $a_1=a_2$ and $a_3$. An ideal graphene flake can be treated as an oblate spheroid with $\alpha=a_3/a_1\rightarrow0$, while CNT can be treated as a prolate spheroid with $\alpha\rightarrow\infty$. This difference in $\alpha$ was theoretically predicted to make graphene much better filler than CNTs [28]. Assuming randomly oriented fillers and incorporating TBR, we can write for MLG composites

$$K = K_p [3K_m + 2f(K_p - K_m)/[(3-f)K_p + K_m f + R_B K_m K_p f / H].$$  (1)

Here $R_B$ is TBR at the graphene/matrix interface, while $K_p$ and $K_m$ are thermal conductivity of the filler and matrix materials, respectively. Graphene has large phonon mean-free path $\Lambda\sim775$ nm at RT [23], which is comparable to $L$. To account for the size effects on heat conduction inside MLG, we altered EMA by introducing $K_p = (1/3)Cv\Lambda_{eff}$, where $1/\Lambda_{eff} = 1/\Lambda + 1/L$, $C$ is the specific heat and $v$ is the phonon velocity. For simplicity, we assumed $K_p/K_m \sim 1000$ for both MLG and CNTs and took TBR values for CNTs and graphene as $\sim8.3\times10^{-8}$ m$^2$ KW$^{-1}$ [17] and $\sim3.7\times10^{-9}$ m$^2$ KW$^{-1}$[29], respectively. Figure 4a shows calculated ratio $K/K_m$ vs. $f$ for MLG ($L=100$ nm) and CNT composites. It confirms that MLG can produce higher TCE than CNTs even as one varies $\alpha$ and diameter $D$ of CNTs in a wide range. One should note that the thermal conductivity model of the graphene fillers is based on the kinetic theory and only considers the lateral size effect without accounting for other effects such as the substrate scattering.

We now use the modified EMA to extract actual TBR in our nanocomposite graphene-MLG-epoxy TIMs by fitting calculated $K$ to the experimental data and varying $R_B$ value (Figure 4b). For MLG, we use $\alpha$=0.01 ($\approx H/L$) and assume $\Lambda$=775 nm [26]. The best match with experiment is attained at $R_B$=3.5×10$^{-9}$ Km$^2$W$^{-1}$. This value is small and consistent with the molecular dynamics (MD) simulations [29]. Our own calculations indicate that for higher $R_B$, TCE does not increase with $f$ linearly but starts to saturate. In addition to the geometrical factors, the reduction of TBR at the filler/matrix interface is another key condition for achieving high TCE for graphene-MGL nanocomposite.

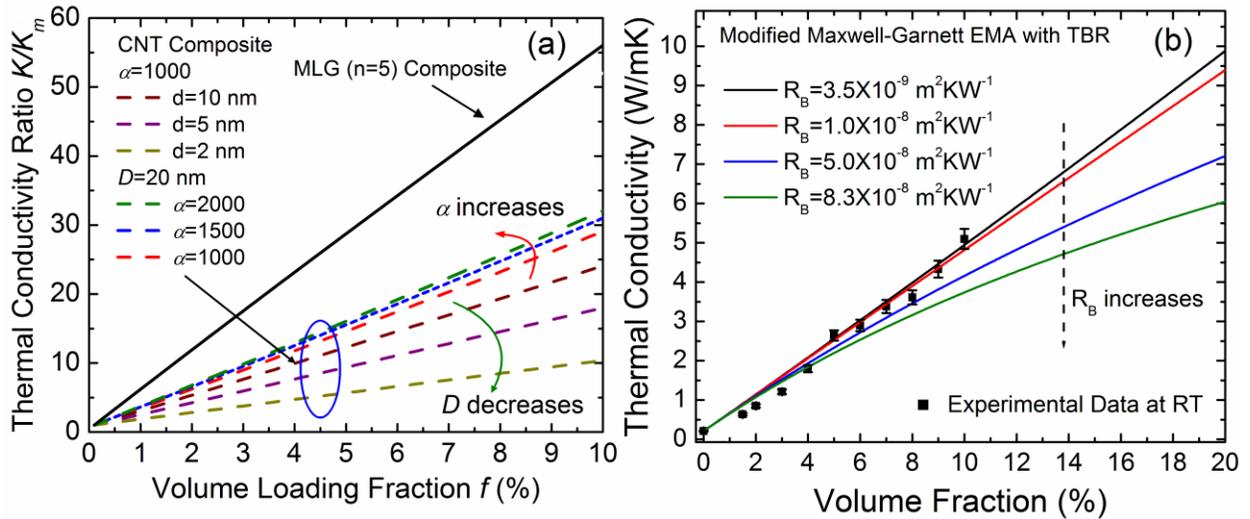

**Figure 4:** Calculated thermal conductivity of graphene-MLG-polymer TIMs. (a) Comparison of thermal conductivity of MGL (*n*=5) and CNTs based TIMs. Note the dependence of thermal conductivity of CNT composites on the aspect ratio and diameter. (b) Thermal conductivity of MLG-polymer TIMs as a function of loading calculated for different values of TBR at the MLG/matrix interface. Fitting of the theoretical curves to the experimental data was used for extraction of the actual values of TBR.

Recent *ab initio* density function theory (DFT) and MD study [30] suggested a possibility of extraordinary $K$ enhancement in ordered graphene composites ($K/K_m \approx 360$ at $f \approx 5\%$) due to graphene's planar geometry and strong coupling of the functionalized graphene to the organic molecules with the corresponding decrease in Kapitza resistance. This implies that certain phonon modes excited in graphene and couple well to those in organic molecules and the mismatch in the phonon DOS between graphene – matrix is smaller than between CNT/matrix. Our experimental results are in line with these DFT and MD predictions [29-30].

It follows from our analysis that graphene's geometry ($\alpha \rightarrow 0$ in graphene as opposed to $\alpha \rightarrow \infty$ in CNTs) and lower Kapitza resistance are the key factors in achieving outstanding TCE. The role of the percolation threshold is not clear yet. Theory suggests that $f_T \sim 1/\alpha$ [19], which explains the low electrical percolation $f_T$ for CNTs. This can also indicate that for MLG, $f_T$ should be much larger and heat conduction is assisted, instead, by better graphene and MLG thermal coupling to the matrix. The latter conclusion is supported by the extracted value of $R_B$ and theoretical estimates of Kapitza resistance. These considerations do not exclude attachments of graphene and MLG flakes to each other with good thermal links without forming a completely percolated network. In the examined $f$ range our TIM samples remained electrically insulating with the measured electrical resistivity of $\rho \approx 1.4 \times 10^9$ $\Omega$-cm. Below we discuss a possibility of the strong $K$ enhancement without a substantial decrease of $\rho$ value in more details.

From the extensive trial-and-error study, we established that it is essential to have both graphene and MGL in the nanocomposite, i.e. to have the graphene-MLG mixture, for achieving maximum TEC. The single-layer or bilayer graphene have greater flexibility to form the thermal links while $K_p$ in MGL ($n>3$) is subject to less degradation due to phonon – boundary scattering [23]. The TIM performance, defined by $R_{TIM}$, depends not only on $K$ but also on $BLT$ [3-5]. We estimated that our samples have $BLT \leq 5$ μm at relevant $P$. $BLT$ evolution with $f$ follows the equations $BLT \sim \tau_y/P$, with the yield stress given as $\tau_y = A[1/((f/f_m)^{1/3}-1)^2)$, where $A$ is a constant and $f_m$ is the maximum filler particle volume fraction [2]. Using an approximate $BLT$ and measured TCE of 2300%, we conservatively estimate that $R_{TIM}$ of the nanocomposite graphene-MLG TIMs should be, at least, on the order of magnitude smaller than that of conventional or CNT based TIMs. The achieved TCE at $f$=10% is higher than that in graphite composites [31], GnP – CNT epoxy composites [32-33], graphite nanocomposites [34], or chemically functionalized graphite composites [35] at the same or even higher carbon loading.

In order to evaluate the effectiveness of graphene-MLG-based TIMs in the practical setting of the two proximate surfaces and TIM between them, we measured the thermal conductivity across the thermal contact. We prepared sandwiches of the two mating surfaces made of aluminum with the TIM in between two surfaces. First, we started with a commercial thermal grease as the TIM material which has Al and $ZnO_2$ particles as the filler materials [36]. The thermal conductivity of the stacked metal-grease-metal sandwiches was measured using the same laser flash technique. The thermal conductivity of the thermal grease determined in our experiments was ~5.8 W/mK, which compares well with the value provided by the vendor. As the next step, we modified the grease by adding a small quantity ($f$=2 vol. %) of our mixture of graphene-MLG, and prepared several sandwiches of the metal-TIM-metal. The thermal conductivity of the total structure was measured again following the same procedure. The extracted thermal conductivity of our graphene-MLG-grease TIM was found to be ~14 W/mK at RT. This corresponds to $K/K_m$ ratio of ~2.4, i.e. TCE factor of ~1.4, at the very small 2% loading fraction.

For comparison, in the case of our graphene-MLG-epoxy composite the TCE factor is ~3 at 2 % loading, which corresponds to $K/K_m$~4 (see Figure 3 (a)). Although the TCE factor in the tested commercial grease with graphene is smaller than that in the graphene-epoxy composite, it is still significant. It is reasonable to assume that in the commercial grease the TCE factor is smaller than in the graphene-MLG-epoxy nanocomposites owing to the presence of other filler particles (Al and $ZnO_2$) with the relatively low intrinsic $K$. A different graphene – matrix coupling can also affect the $K$ value. In the graphene-epoxy composites we also start with the much smaller matrix thermal conductivity $K_m$. A hybrid mixture of graphene-MLG and Al and $ZnO_2$ can be an efficient filler owing to a complex interactions among different filler particles [6].

It is important to note that by using ~2% loading fraction, we kept the viscosity and other important mechanical characteristics, such as conformity and spreadability, of the original thermal grease unaffected. Conformability allows TIM to fill the microscopic valleys on the surface of the mating surfaces, thus displacing air, which is thermally insulating. The spreadability allows one to minimize the TIM thickness: the larger thickness would result in the higher thermal resistance. Table I summarizes previously reported thermal conductivity values for TIMs with various fillers. The data shows that the measured TCE in our graphene-MLG-epoxy composites is indeed extremely high. Implementing this industry testing protocol with the commercial thermal grease, we have demonstrated that graphene-MLG nanocomposite prepared under the optimum conditions are promising as the next generation TIMs.

Often, the large TCE factors are accompanied by increasing electrical conductivity $\sigma$, i.e. decreasing $\rho$ (see references in Table I). In our case, we observed a record-high TEC without a substantial change in $\rho$ in the examined $f$ range. The increase of TEC without decreasing $\rho$ was reported in a few other studies. For example, a substantial enhancement of $K$ in the composites with CNTs at 1 wt. % loading was reported in Ref. [45]. The electrical conductivity $\sigma$ of the composites remained low $10^{-11} - 10^{-9}$ Scm$^{-1}$ in these samples. The low $\sigma$ of $10^{-15} - 10^{-9}$ Scm$^{-1}$ in the SWNT/PS composites with enhanced $K$ was also reported [46]. Moreover, there were studies where the increasing $K$ was accompanied by the decreasing $\sigma$ in the composite with the same filler fraction [32].

The increase in $K$ without substantial change in $\sigma$, observed in our experiments, can be explained by the following. The strong increase in the electrical conductivity in the composite with the electrically insulating matrix requires formation of the percolation network. In our case, we have enhancement of $K$ owing to the present of graphene-MLG fillers, perhaps with partial ordering, while the complete percolation network is not formed. The heat can be conducted through the matrix while the electrical current cannot. The narrow layers of the epoxy matrix may not present a substantial thermal resistance while blocking the electric current. According to the theory [19, 47], an increase in the thickness of the polymer layer from zero to 10 nm does not affect significantly the heat transport while such an increase in the width of the tunneling barrier for the electrons would effectively eliminate the electrical transport.

In conclusion, we synthesized graphene-MLG nanocomposite polymer TIMs and demonstrated the extremely high TCE factors at low filler loadings. The TIM testing has been conducted in the industry-type settings ensuring that all other TIM characteristics are in acceptable range for TIMs' practical applications. The TCE of 2300% at $f$=10% loading is higher than anything reported to-date. We explained the unusual enhancement by (i) high intrinsic $K_i$ of graphene and MLG, (ii) low Kapitza resistance at the graphene/matrix interface; (iii) geometrical shape of graphene/MLG flakes ($a \rightarrow 0$); (iv) high flexibility if MLG ($n$<5); and (v) optimum mix of graphene and MGL with different thickness and lateral size. Additional benefits of the graphene-based composites, which come at now additional expense, are their low coefficient of thermal expansion [23] and increased mechanical strength [24-25]. We have also demonstrated a possibility of achieving $K$~14 W/mK in the commercial thermal grease via addition of only $f$=2% of the optimized graphene-MLG nanocomposite mixture. The graphene-based TIMs have thermal resistance $R_{TIM}$ reduced by orders-of-magnitude and be can produced inexpensively on an industrial scale, thus, allowing for the first graphene application, which consumes this material in large quantities.

## Methods: Synthesis of the Graphene-MLG Nanocomposite Thermal Interface Materials

The graphene-MLG nanocomposites were prepared by ultrasonication (~10-12 hrs) of natural graphite in aqueous solution of sodium cholate. The solution was left for ~1 hr to settle followed by removal of thick graphite flakes. The ultrasonicated solution underwent sedimentation processing in a centrifuge. The centrifugation was performed at 15 K-rpm for 5 min. After centrifugation the top layer was decanted and dried in a vacuum oven. It was again dispersed in water by the high-sheer mixing followed by ultrasonication for ~2 hrs. The solvent was dried at 60°C in a vacuum oven leaving graphene and MLG consisting of 1-10 stacked atomic monolayers. The epoxy resin (diglycidyl ether of bisphenol F, EPON 862, Hexion) was added to the suspension following an in-house developed procedure (see Supplemental Information). The curing agent (diethyltoluenediamine, EPI-CURE) was added under continuous stirring in a ratio of epoxy to curing agent of 100:26 by weight. The homogeneous mixture of epoxy and graphene-MLG nanocomposite was loaded into a custom stainless steel mold, heated and degassed in vacuum for curing. The composites were cured at 100 °C for 2 h and at 150 °C for additional 2 h to complete the curing cycle. A large number of samples were prepared with different graphene loadings varying between 1-10% of volume fraction.

## *Acknowledgements*

This work was supported, in part, by the Office of Naval Research (ONR) through award N00014-10-1-0224 on graphene heat spreaders and by the Semiconductor Research Corporation (SRC) and Defense Advanced Research Project Agency (DARPA) through FCRP Center on Functional Engineered Nano Architectonics (FENA). The authors acknowledge useful discussions on TIMs with Intel Corporation engineers, and thank the former members of the Nano-Device Laboratory (NDL) – Dr. S. Ghosh (Intel), Dr. D. Teweldebrhan (Intel) and Dr. V. Goyal (Texas Instrument) – for their help with thermal measurements. AAB is indebted to Dr. A. Gowda (GE Global Research) for invitation to serve in the Panel on CNT/Graphene TIMs at the ITherm-2010 conference, which provided an extra stimulus to this study.

**References:**

[1] Balandin, *IEEE Spectrum*, **29**, 35-39 (2009).

[2] Garimella, S.V., Fleischer, A.S., Murthy, J.Y., Keshavarzi, A., Prasher, R., Patel, C., Bhavnani, S.H., Venkatasubramanian, R., Mahajan, R., Joshi, Y., Sammakia, B., Myers, B. A., Chorosinski, L., Baelmans, M., Sathyamurthy, P., & Raad, P. E. *IEEE Transactions on Components and Packaging Technologies*, **31**, 801 − 815 (2008).

[3] Prasher, R. *Proceedings of IEEE*, **94**, 1571 − 1585 (2006).

[4] Sarvar, F., Whalley D.C., & Conway, P.P. *Proceeds. Electronics System Integration Technology Conference* (IEEE 1-4244-0553), **2**, 1292-1302 (2006).

[5] Prasher, R. S., Chang, J.-Y., Sauciuc, I., Narasimhan, S., Chau, D., Chrysler, G., Myers, A., Prstic, S. & Hu, C., *Intel Technology Journal*, **9**, 285 − 296 (2005).

[6] Felba, J. Thermally conductive nanocomposites, in *Nano-Bio-Electronic, Photonic and MEMS Packaging* (Springer Science, 2010; DOI 10.1007/978), Editors C.P. Wong, K.-S. Moon and Y. Li, 277 − 314 (2010).

[7] Novoselov, K. S., Geim, A. K., Morozov, S. V.,  Jiang, D., Zhang, Y., Dubonos, S. V., Grigorieva,  I. V.  & Firsov, A. A. *Science*, **306**, 5696, 666-669 (2004).

[8] Hernandez, Y., Nicolosi, V., Lotya, M., Blighe, F. M., Sun, Z., De, S., McGovern, I. T., Holland, B., Byrne, M., GunKo, Y. K., Boland, J. J., Niraj, P., Duesberg, G., Krishnamurthy, S., Goodhue, R., Hutchison, J., Scardaci, V., Ferrari, A. C. &  Coleman, J. N. *Nature Nanotechnology* **3***,* 563 - 568 (2008).

[9] Green, A. A.  & Hersam, M. C., *Nano Lett.* **9**, 12, 4031–4036 (2009).

[10] Lotya, M., Hernandez, Y., King, P. J., Smith, R. J., Nicolosi, V. Karlsson, L. S. Blighe, F. M., De, S., Wang, Z., McGovern, I. T., Duesberg, G. S.  & Coleman, J. N., *J. Am. Chem. Soc.*, **131** (10), 3611–3620 (2009).

[11] Kim, P., Shi, L., Majumdar, A. & McEuen, P. L.  *Phys. Rev. Lett*. **87**, 215502-4 (2001).

[12] Pop, E., Mann, D., Wang, Q., Goodson, K. & Dai, H.  *Nano Lett*. **6**, 1, 96-100 (2006).

[13] Yu, A., Itkis, M. E., Bekyarova E.  & Haddon, R. C. *Appl. Phys. Lett.*, **89**, 133102 (2006).

[14] Bonnet, P., Sireude, D., Garnier, B. & Chauvet, O., *Appl. Phys. Lett.* **91**, 201910 (2007).

[15] Choi, S.U.S., Zhang, Z.G., Yu, W., Lockwood, F.E. & Grulke, E.A., *Appl. Phys. Lett*., **79**, 2252 (2001).

[16] Moisala, A., Lia, Q., Kinlocha, I. A. & Windle, A.H., *Compos. Sci. Technol.* **66**, 10, 1285-1288 (2006).

[17] Huxtable, S., Cahill, D. G., Shenogin, S., Xue, L., Ozisik, R., Barone, P., Usrey, M., Strano, M. S., Siddons, G., Shim, M. & Keblinski, P., *Nat. Mater*. **2**, 731 - 734 (2003).

[18] Kapitza. P. L., *J. Phys. USSR*, **4**,181(1941).

[19] Shenogina, N., Shenogin, S., Xue, L., & Keblinski, P., *Appl. Phys. Lett.*, **87**, 133106 (2005).

[20] Balandin, A. A., Ghosh, S., Bao, W., Calizo, I., Teweldebrhan, D., Miao F. & Lau, C. N., *Nano Lett.* **8**, 3, 902–907 (2008).

[21] Nika, D.L., Pokatilov, E. P., Askerov A. S. & Balandin, A.A., *Phys. Rev. B* **79**, 155413 (2009).

[22] Ghosh, S., Bao, W., Nika, D. L., Subrina, S., Pokatilov, E. P., Lau, C. N. & Balandin, A. A., *Nature Mat.* **9**, 555-558 (2010).

[23] Balandin, A.A, *Nature Mat.*, **10**, 569 - 581 (2011).

[24] Stankovich, S., Dikin, D. A., Dommett, G H. B., Kohlhaas, K M., Zimney, E J., Stach, E A., Piner, R D., Nguyen, S T. & Ruoff, R. S. *Nature* **442**, 282-286 (2006).

[25] Li, D., Müller, M. B., Gilje, S, Kaner R. B. & Wallace, G. G., *Nature Nanotech*. **3**, 101 - 105 (2008).

[26] Calizo, I., Bejenari, I., Rahman, M., Liu G. & Balandin, A.A., *J. Appl. Phys.* **106**, 043509 (2009).

[27] Nan, C.W., Birringer, R., Clarke, D. R. & Gleiter, H., *J. Appl. Phys.* **81**, 6692 (1997).

[28] Xie, S. H., Liu, Y. Y. & Li, J. Y., *Appl. Phys. Lett*. **92**, 243121 (2008)

[29] Konatham, D. & Striolo, A., *Appl. Phys. Lett*, **95**, 163105 (2009).

[30] Konatham, D., Bui, K. N. D., Papavassiliou & D. V., Striolo, A., *Molecular Physics* **109**, 1, 97 – 111 (2011).

[31] Debelak, B. & Lafdi, K., Carbon **45**, 1727 (2007).

[32] Yu, A., Ramesh, P., Sun, X., Bekyarova, E., Itkis, M.E. & Haddon, R.C., *Adv. Mater*. **20**, 4740 (2008).

[33] Lin, C. & Chung, D.D.L., *Carbon*, **47**, 295 (2009).

[34] Fukushima, H., Drzal, L.T., Rook, B.P., Rich & M.J, *J. Thermal Analysis and Calorimetry*, **85**, 235 (2006).

[35] Gangui, S., Roy, A.K. & Anderson, D.P., *Carbon*, **46**, 806 (2008).

[36] Narumanchi, S., Mihalic, M., Kelly, K. & Eesley, G., NREL/CP-540-42972 Conference (2008).

[37] Hung, M. T., Choi, O., Ju, Y. S. & Hahn, H. T., *Appl. Phys. Lett*. **89**, 023117(2006).

[38] Biercuk, M.J., Llaguno, M. C., Radosavljevic, M., Hyun, J. K., Johnson A. T. & Fischer, E., J., *Appl. Phys. Lett*., **80**, 2767 (2002).

[39] Liu, C.H., Huang, H., Wu, Y. & Fan, S. S., *Appl. Phys. Lett*., **84**, 4248 (2004).

[40] Yu, A., Ramesh, P., Itkis, M. E., Bekyarova, E. & Haddon, R. C., *J. Phys. Chem. Lett*., **111**, 7565 (2007).

[41] Yu, W., Xie, H. & Chen, W., *J. Appl. Phys*., **107**, 094317 (2010).

[42] Nikkeshi, S., Kudo, M. & Masuko, T., *J. Appl. Pol. Science*, **69**, 2593 (1998).

[43] Ohashi, M., Kawakami, S., Yokogawa, Y. & Lai G. C., *J. Amer. Ceramic Soc*., **88**, 2615 (2005).

[44] Hodgin, M.J. & Estes, R.H., Proceeds. of the National Electronic Packaging and Production Conference, 359 (1999).

[45] Jakubinek, M.B., White, M. A., Mu M. & Winey, K. I., *Appl. Phys. Lett*., **96**, 083105 (2010).

[46] Tchoul, M.N., Ford, W.T., Ha, M. L. P., Sumarriva, I. C., Grady, B. P., Lolli, G., Resasco, D. E. & Arepalli S., *Chem. Mater*., **20**, 3120 (2008).

[47] Shenogin, S., Xue, L. P., Ozisik, R., Keblinski, P. & Cahill, D. G., *J. Appl. Phys.*, **95**, 8136 (2004).

**Table I: Thermal Conductivity Enhancement in TIM Composites**

| Filler | TCE | Fraction | Base Material | Method | Reference |
|---|---|---|---|---|---|
| MWNT | 150 % | 1.0 vol. % | oil | transient hot wire | [15] |
| SWNT | 125 % | 1.0 wt. % | epoxy | comparative method | [38] |
| p-SWNT | 350 % | 9.0 wt. % | epoxy | laser flash | [13] |
| CNT | 65 % | 3.8 wt. % | silicone | ASTM | [39] |
| GNP | 3000 % | 25.0 vol. % | epoxy | laser flash | [40] |
| GON | 30% - 80% | 5.0 vol. % | glycol and paraffin | comparative method | [41] |
| SWNT | 55 % | 7.0 wt. % | PMMA | guarded plate | [14] |
| GNP | 10 % | 1.0 vol. % | epoxy | transient hot wire | [37] |
| Ni | 566 % | <30 % | epoxy | laser flash | [42] |
| ALN | 1900 % | 60 % | epoxy | ASTM | [43] |
| BN | 650 % | 30 wt. % | epoxy | ASTM | [44] |
| SWNT | 50 % | 1 wt. % | polystyrene | steady state method | [45] |
| Graphite | 1800 % | 20 wt. % | epoxy | laser flash | [31] |
| Graphene - MLG | 2300 % | 10 vol. % | epoxy | laser flash | This work |